# How do black holes move, as quantum objects or as classical objects?

## C. L. Herzenberg

**Abstract**
*Results of a recent study of the transition between quantum and classical behavior are applied to black holes. The study led to a criterion separating quantum from classical behavior on the basis of mass or size, dependent on local effects of cosmic expansion. Application of this criterion to black holes indicates that the motion of smaller black holes will be characteristically quantum mechanical, while the motion of larger black holes must be classical, with a threshold distinguishing these behaviors at a Schwartzschild radius of roughly the size of a nucleon.*

**Keywords**: black hole, quantum-classical transition, critical length, Schwartzschild radius, evaporation time, Hawking radiation, Hubble expansion, Hubble time, cosmic expansion, quantum behavior

**Introduction**

Since the early 1970s, a great deal of work has been done on the quantum aspects of black hole physics, with important advances in topics such as Hawking radiation and associated black hole evaporation, black hole entropy, and various related topics.[1,2] However, a fairly simply stated and straightforward question may still be asked, and that is the question of how the position in space and the motion of a black hole should best be described, by using quantum mechanics or by using classical mechanics.

Here we are not concerned directly with the internal dynamics of a black hole, only the behavior of a black hole as an entire object at rest or in motion. Under what circumstances is the translational motion of a black hole best described by the classical equations of motion, and under what circumstances is it best described by a quantum mechanical treatmenr of a free object at rest or engaged in constant velocity motion? While aspects of related questions have been addressed, such as the general question of the decoherence of quantum states in black holes, a straightforward approach to this question seems worth examining.[2]

So, how do black holes move – as a classical objects, or as quantum objects? Or what? Under what circumstances can a black hole behave quantum mechanically as an entire object? In this paper we seek to address in a straightforward manner this basic behavior of black holes, and attempt to determine under what circumstances the motion of a black hole will be properly described quantum mechanically and under what circumstances a black hole will move classically.

But, more generally, what distinguishes classical from quantum objects? The study of decoherence effects in matter has addressed many but not all issues.[3] In the past, in the absence of a crisp criterion to distinguish between quantum and classical, an identification of classical with macroscopic has often been tentatively and somewhat arbitrarily accepted.[4] Recent research results not only support this categorization but provide a fundamental reason for its existence and in addition provide a specific size criterion that distinguishes ranges of quantum and classical behaviors.[5-8] This research leads to the conclusion that the motion of sufficiently small objects will be quantum mechanical, while the motion of large objects of necessity must be classical, as a consequence of a specific cause, the local effects of cosmic expansion. A realistic threshold length separating quantum from classical behavior in ordinary matter emerges from the local effects of Hubble expansion on matter. We now examine the possible implications of extending these results to black holes.

**Background – black holes**

The concept of a black hole was developed in a classical context. General relativity, a classical theory of gravity, describes a black hole in terms of a region of empty space with a pointlike singularity at the center and an event horizon at the outer edge. However, practically speaking, the center of a black hole could be regarded not so much as a singularity but rather as a mass compressed into the smallest possible volume. In classical gravitational theory, black holes are present as soliton-like solutions of the gravitational equations.[9] Black holes are characterized by their position in space as well as by possible different internal states.[9]

Our understanding of black holes must change to some extent when quantum mechanics is taken into account. While a satisfactory quantum theory of black holes has not yet been developed, one can take as a working assumption that when quantum mechanics is applied to black holes, and a black hole is considered as a quantum object, its state would be described by a wave function.[9] Such a wave function would contain information about the motion of the black hole in or through the surrounding space-time, and also information about internal states of black hole excitations. For the simplest case of free motion of a quantum black hole in an asymptotically flat space-time, one might expect the overall wave function to be the product of two factors, an exponential involving the 4-momentum of the black hole and its center-of-mass coordinates describing its external motion, with an internal wave function describing internal states of the black hole.[9] Quantum states involving an external wave function describing a wave packet or superposed waves might then presumably be of use to describe black holes in other than a quantum state of pure constant velocity motion. Most attention to quantum effects in the literature has been addressed to internal excitation states of a black hole and related emissions, but here we can direct our attention primarily to the quantum behavior associated with the motion of the entire black hole.

In classical general relativity, a black hole can have any mass, including an arbitrarily small mass, since any quantity of matter that is sufficiently compressed could become a black hole.[10] However, when black holes form in accordance with known natural processes, only a few mass ranges are regarded as realistic. These include supermassive black holes containing millions to billions of times the mass of the sun that are expected to exist in the center of most galaxies; intermediate-mass black holes, with masses measured in thousands of solar masses; and stellar-mass black holes, formed by the collapse of individual stars.[1, 10] All contemporary black hole formation is thought to take place through gravitational collapse, with the smallest mass that can collapse and form a black hole in this process being comparable to the mass of the sun.

Sub-solar-mass black holes have received somewhat less attention than more massive black holes. It appears that no known process currently active in the universe can form black holes of less than stellar mass, so that black holes of mass less than approximately the mass of the sun are thought not to be able to be formed by natural processes in the present universe.[10, 11] There are a few ways in which such smaller black holes might be formed, or might have been formed in the past. These include formation by evaporation (emission of thermal quanta) of larger black holes; however, if the initial mass of the black hole were a stellar mass, the time required for it to lose most of its mass by emission of Hawking radiation would be much longer than the age of the universe, so that any contemporary small black holes are not expected to have formed by this method. Primordial black holes might have been formed during the Big Bang, which would have produced sufficiently high pressures to create smaller black holes. However, while only primordial black holes could be present naturally as small black holes, it appears that primordial black holes remain to be detected.[10, 11] Another interesting source of smaller black holes might be the possibility of production of micro black holes in accelerator experiments.[12]

The huge range of masses between stellar dimensions and the Planck mass seems to a large extent to have been neglected in the literature. Posing the question of, over what mass range or size range might black hole behavior be dominated by quantum mechanics, may be of interest here.

**Background – quantum-classical transition**

What distinguishes quantum from classical objects? What is the structure of the transition from quantum to classical – is it smooth or abrupt? Where is the borderline between the applicability of the two kinematic concepts? Such questions have been raised and to a limited extent addressed, mainly in connection with the concept of the appearance of classical behavior in quantum theory through decoherence with the environment.[3] Recently, a new approach has been taken toward explaining within quantum mechanics the classical appearance of our macroscopic world, and evidence has been found that a transition from quantum behavior to classical behavior will occur in all ordinary matter as a result of the presence of Hubble expansion within extended objects.[5, 6, 8] Quantum behavior was found to occur for sufficiently small objects, while large objects were, because of their size, constrained to behave classically, with a transitional range of behaviors in between. In effect, free particle quantum states of extended objects appeared to be forced into states described by localized wave packets by the presence within the objects of Hubble expansion velocities.[5, 6, 8] In addition, a conceptually completely independent derivation based on stochastic quantum mechanics led to similar results.[7]

Criteria based on mass or size were derived that distinguished classical from quantum behavior in matter.[5 – 8]

An approximate threshold size separating quantum from classical behavior in objects due to this effect has been evaluated in terms of a critical length $L_{cr}$. The critical length, or threshold length, is given by the expression:[5 – 8]

$$L_{cr} = [h/(4\pi M H_o)]^{1/2} \quad (1)$$

Here, M is the mass of the object, $H_o$ is the Hubble constant, and h is Planck's constant. For a given value of the mass, an object with a linear size greater than this value would be expected to behave classically, while an object whose linear dimensions are considerably less would be expected to behave quantum mechanically as a whole object, with intermediate behavior at around this size.

**Quantum-classical transition in black holes**

While the expression for the critical length was originally derived in the context of quantum and classical behavior of ordinary matter, it may be of interest to examine what could be learned by a tentative application to black holes.

We will take the Schwartzschild radius of a black hole as a measure of its linear size. The Schwartzschild radius for a black hole of mass M is given by:[2, 10]

$$R_S = 2GM/c^2 \quad (2)$$

Here, G is the gravitational constant and c is the velocity of light.

Let us determine under what conditions the linear size, or Schwartzschild radius, of a black hole would be equal to the critical length, by setting these parameters equal to each other:

$$R_S = L_{cr} \quad (3)$$

Since both the value of the critical length and the value of the Schwartzschild radius depend on the mass of the object, the requirement imposed by Eqn. (3) determines a threshold mass that will distinguish smaller black holes that might be expected to behave quantum mechanically from larger black holes that might be expected to behave classically, if these considerations do indeed apply.

If we evaluate the threshold mass by inserting Eqn. (1) and Eqn. (2) into Eqn. (3), we find that this threshold mass for a black hole must be equal to:

$$M_{th} = [hc^4/(16\pi G^2 H_o)]^{1/3} \quad (4)$$

The threshold mass value that may be expected to separate predominantly classical black holes from black holes that exhibit explicit quantum behavior is thus found to be determined by natural constants.

If we evaluate Eqn. (4) numerically, we find that the numerical value of the threshold mass that would separate smaller mass black holes that presumptively would behave quantum mechanically from larger mass black holes that would behave classically turns out to be roughly $2 \times 10^{12}$ kilograms. If such a black hole had been formed by gravitational collapse, it would have corresponded to an object of ordinary density and roughly a kilometer on a side prior to gravitational collapse.

Perhaps more interesting is the Schwartzschild radius associated with such a threshold-sized black hole. (We limit our attention to non-rotating black holes for which the event horizon is identical with the Schwartzschild radius.) If we evaluate the Schwartzschild radius of such a black hole of threshold size by combining Eqn. (4) with Eqn. (2), we find:

$$R_{Sth} = [hG/(2\pi c^2 H_o)]^{1/3} \qquad (5)$$

Thus, the threshold Schwartzschild radius of a black hole of mass corresponding to the quantum-classical transition is also determined in terms of physical constants.

If we evaluate numerically the threshold Schwartzschild radius for a black hole, we find that it is approximately $3 \times 10^{-15}$ meters. This is very roughly comparable to the size of a nucleon. Thus, it appears that black holes smaller than about the size of a nucleon might be expected to behave quantum mechanically as entire objects, while black holes larger than approximately the size of a nucleon might be expected to behave classically, on the basis of these considerations.

**Evaporation of black holes**

Black holes appear to have the capability of losing mass-energy by the Hawking radiation process that takes place by polarization of the vacuum and which can be treated by describing the black hole's gravitational field by classical general relativity while the surrounding vacuum space-time is described by quantum field theory.[13] When black holes lose energy by Hawking radiation (by emission of neutrinos, photons, gravitons, etc.), the power can be calculated, and the thermal emission turns out to be small for large black holes and large for small black holes, as it depends inversely on the square of the mass. An evaporation time can be calculated from the emitted power under the assumption that the amount of any simultaneous infalling matter or radiation is negligible. This lifetime depends on the cube of the initial mass of the black hole and is temperature dependent in that it reflects the number of degrees of freedom in the emission process.[12] An estimate of this evaporation time is given by the equation:[14]

$$t_{ev} = 640\pi^2 G^2 M_o^3/hc^4 \qquad (6)$$

where $M_o$ is the initial mass of the black hole undergoing evaporation by the emission of mass-energy.

Next, we can evaluate the evaporation time or lifetime of a black hole having the threshold mass. Inserting the expression for the threshold mass from Eqn. (4) into Eqn. (6) we find for the evaporation time of a black hole having an initial mass equal to the threshold mass:

$$t_{evth} = 40\pi/H_o \qquad (7)$$

Thus, the evaporation time of a black hole of threshold mass is directly related to the Hubble constant, and thus to the Hubble time. Using a value of the Hubble constant of $2.3 \times 10^{-18}$ per second, we find for the evaporation time a value of roughly $10^{12}$ years, somewhat longer than the time since the Big Bang.

Somewhat differing estimates of evaporation lifetimes exist in the literature. It is pointed out that a black hole with a mass of the order of $10^{15}$ grams (the mass of a typical asteroid) and corresponding Schwartzchild radius of the order of $10^{-13}$ cm. (about the size of a nucleon) will have an evaporation lifetime of the order of the age of out universe.[1]

A related estimate of the evaporation time for black holes from the literature gives:[15]

$$t_{ev} \approx 10^{10} \text{ years } (M_o/10^{15} \text{ grams})^3 \qquad (8)$$

This tells us that only those black holes with masses less than or comparable to $10^{15}$ grams have the evaporation times shorter than the age of the universe. Use of this estimate for a threshold mass $2 \times 10^{12}$ kilograms would lead to an evaporation time estimate of 80 billion years, longer than but within an order of magnitude of the lifetime of the visible universe.

Thus, depending on the estimates used, the threshold mass for a black hole may be only moderately above the mass of a black hole having an evaporation time comparable to the time since the Big Bang.

**Discussion**

Thus, we find that black holes of the threshold mass have evaporation lifetimes very roughly in the range of the time since the big bang, within a couple of orders of magnitude, and, if anything, have longer lifetimes. Thus there is a possibility that primordial black holes of the threshold mass might still exist.

That a black hole with the threshold mass turns out to have an evaporation lifetime roughly in the range of the lifetime of the universe is of course not just a coincidence. We can examine this further by taking the estimate for the evaporative lifetime of a black hole given by Eqn. (6) and equating it to the lifetime of the universe to determine the magnitude of the mass of a black hole which would barely have survived since the Big

Bang. We will use the inverse of the Hubble constant as an approximation for the lifetime of the universe. If we do this, the criterion becomes:

$$H_o^{-1} = 640\pi^2 G^2 M_o^3/(hc^4) \tag{9}$$

From this we can evaluate the mass of a black hole that would have an evaporation lifetime equal to the inverse of the Hubble constant as:

$$M_{oH-1} = [hc^4/(640\pi^2 G^2 H_o)]^{1/3} \tag{10}$$

When we compare this with the threshold mass that would distinguish quantum from classical behavior in black holes derived in Eqn. (4), we find that the expressions are identical apart from numerical factors, and correspond to masses that are within about a factor of 5 of each other.

Why is it that these two quantities that are associated with rather different properties of a black hole are so similar? It would appear that there may be an underlying unity here, based on the fact that both parameters address fundamental quantum aspects of black holes in our present universe. While non-classical black hole quantum states can be constructed by using the superposition principle, it has been reported that most such states decohere through their own Hawking radiation.[2] This would seem to present a somewhat puzzling possible connection.

As just discussed, it is usually accepted that black holes have finite lifetimes due to the emission of Hawking radiation. However, the existence of Hawking radiation has been called into question by some authors, for several reasons. In Hawking's model of black hole radiation, quantum field theory is superimposed on curved space-time, with gravity described classically according to general relativity; this is a semiclassical approach in which only the matter fields are quantized and black hole radiation is driven by quantum fluctuations of these fields.[14] More important, so far there appears to be no convincing observational evidence of Hawking radiation.[1, 14] If Hawking radiation does not exist, the evaporation mechanism that has been thought to destroy black holes, especially small black holes, would not be present, and thus these quantum black holes might be stable instead of decaying. There thus remains the possibility that black holes having masses or radii below the threshold values might be intrinsically stable objects, with their Hawking radiation suppressed by other quantum effects, perhaps in analogy with the quantum states of an electron which classically would be constantly accelerating round an atom but does not radiate despite the predictions of classical electrodynamics.[16] Thus, it is conceivable that gravitationally dominated black-hole-like structures might still exist at these lower masses, with the emission of Hawking radiation suppressed. If that were the case, many more primordial black holes might still exist in our present universe, and these might include black holes whose motion is best described by quantum mechanics rather than classical mechanics.

Is the present approach truly applicable to black holes? We don't know, but at the very least it provides new ideas and a general criterion for us to work with in attempting to clarify distinctions between quantum and classical behavior in black holes.

**References**

1. Gorini, V., G. Magli, and U. Moschella, "The physics of black holes (an overview)," in *Classical and Quantum Black Holes*, eds. P. Fré, V. Gorini, G. Magli, and U. Moschella (Institute of Physics Publishing, Bristol, 1999).

2. Kiefer, C. "Thermodynamics of black holes and Hawking radiation," in *Classical and Quantum Black Holes*, eds. P. Fré, V. Gorini, G. Magli, and U. Moschella (Institute of Physics Publishing, Bristol, 1999).

3. Joos, E., "Introduction," in *Decoherence and the Appearance of a Classical World in Quantum Theory*, eds. D. Giulini, E. Joos, C. Kiefer, J. Kupsch, I.-O. Stamatescu and H. D. Zeh (Springer, Berlin, 1996).

4. Zurek, W. H., "The environment, decoherence, and the transition from quantum to classical," in *Quantum Gravity and Cosmology*, eds. J. Pérez-Mercader, J. Solà, and E. Verdaguer (World Scientific, Singapore, 1992).

5. Herzenberg, C. L., "Becoming classical: A possible cosmological influence on the quantum-classical transition," *Physics Essays* **19**, No. 4 (December 2006).

6. Herzenberg, C. L., "Why our human-sized world behaves classically, not quantum-mechanically: A popular non-technical exposition of a new idea," arXiv:physics/0701155, http://arxiv.org/abs/physics/0701155 (January, 2007).

7. Herzenberg, C. L., "The role of Hubble time in the quantum-classical transition," *Physics Essays* **20**, No. 1 (March 2007).

8. Herzenberg, C. L., "Is there a cosmological influence on the quantum-classical transition?" *Bulletin of the American Physical Society* **52**, No. 3, p. 62 (April 2007).

9. Frolov, V.P. and I.G. Novikov, *Black Hole Physics: Basic Concepts and New Developments* (Kluwer Academic Publishers, Dordrecht, The Netherlands, 1998).

10. "Black hole," Wikipedia, http://en.wikipedia.org/wiki/Black_hole (accessed 20 August 2007).

11. Novikov, I. G. and V. P. Frolov, *Physics of Black Holes* (Kluwer Academic Publishers, Dordrecht, The Netherlands, 1989).

12. Coyne, D. G., "The experimental quest for primordial black holes," in *International Symposium on Black Holes, Membranes, Wormholes, and Superstrings*, eds. S. Kalara and D. V. Nanopoulos (World Scientific, Singapore, 1993).

13. Luminet, J.-P., "Black Holes: A General Introduction," in *Black Holes: Theory and Observation*, eds. F. W. Hehl, C. Kiefer, and R. J. K. Metzler (Springer-Verlag, Berlin, 1998).

14. Rabinowitz, M., "Black Hole Paradoxes," in *Trends in Black Hole Research*, ed. P. V. Kreitler (Nova Science Publishers Inc., New York, 2006).

15. Shapiro, S. L. and S. A. Teukolsky, *Black Holes, White Dwarfs, and Neutron Stars: The Physics of Compact Objects* (John Wiley & Sons, New York, 1983).

16. "Micro black hole," Wikipedia, http://en.wikipedia.org/wiki/Micro_black_hole (accessed August 20 2007).

C. L. Herzenberg

carol@herzenberg.net

Chicago, IL USA